\documentclass[12pt]{article}
\usepackage[centertags]{amsmath}
\usepackage{amsfonts}
\usepackage{amssymb}
\usepackage{amsthm}
\usepackage{newlfont}
\usepackage{epsfig}
\usepackage{amscd}

\newcommand{\NN}{{\mathbb N}}

\newcommand{\RR}{{\mathbb R}}

\newcommand{\beq}{\begin{equation}}
\newcommand{\eeq}{\end{equation}}
\newcommand{\ba}{\begin{array}}
\newcommand{\ea}{\end{array}}
\newcommand{\bea}{\begin{eqnarray}}
\newcommand{\eea}{\end{eqnarray}}

\begin{document}

\begin{center}
{\large \sc \bf The general solution of the matrix equation \\
$w_t+\sum\limits_{k=1}^nw_{x_k}\rho^{(k)}(w)=\rho(w)+[w,T\tilde\rho(w)]$}

\vskip 15pt

{\large  P. M. Santini$^{1,\S}$ and A. I. Zenchuk$^{2,\S}$}

\vskip 8pt

{\it
$^1$ Dipartimento di Fisica, Universit\`a di Roma "La Sapienza" and\\
Istituto Nazionale di Fisica Nucleare, Sezione di Roma 1\\
Piazz.le Aldo Moro 2, I-00185 Roma, Italy}

\smallskip

{\it $^2$ Center of Nonlinear Studies of the Landau Institute for
Theoretical Physics \\
(International Institute of Nonlinear Science)\\  Kosygina 2,
Moscow, Russia 119334}

\vskip 5pt

$^{\S}$e-mail:  {\tt paolo.santini@roma1.infn.it, zenchuk@itp.ac.ru}

\vskip 5pt

{\today}

\end{center}

\begin{abstract}

\noindent We construct the general solution of the equation
$w_t+\sum\limits_{k=1}^nw_{x_k}\rho^{(k)}(w)=\rho(w)+[w,T\tilde\rho(w)]$,
for the $N\times N$ matrix $w$, where $T$ is any constant diagonal
matrix, $n,~N\in\NN_+$ and $\rho^{(k)},~\rho,~\tilde\rho:\RR\to\RR$
are arbitrary analytic functions. Such a solution is based on the
observation that, as $w$ evolves according to the above equation,
the evolution of its spectrum decouples, and it is ruled by the
scalar analogue of the above equation. Therefore the eigenvalues of
$w$ and suitably normalized eigenvectors are the $N^2$ Riemann
invariants. We also obtain, in the case $\rho=\tilde\rho=0$, a
system of $N^2$ non-differential equations characterizing such a
general solution. We finally discuss reductions of the above matrix
equation to systems of $N$ equations admitting, as Riemann
invariants, the eigenvalues of $w$. The simplest example of such
reductions is a particular case of the gas dynamics equations

\end{abstract}

\section{Introduction}

It is well-known that the general solution of the scalar first order
quasi-linear Partial Differential Equation (PDE): \beq \label{equ-u}
u_t+\sum\limits_{k=1}^nu_{x_k}\rho^{(k)}(u)=\rho(u),~~u:\RR^{n+1}\to\RR,
\eeq can be constructed by the method of characteristics
\cite{Whitham}, converting (\ref{equ-u}) into the system of ODEs:
\beq \label{charact1}
 \ba{l}
\frac{du}{dt}=\rho(u), \\
\frac{dx_k}{dt}=\rho^{(k)}(u)~,k=1,..,n,
\ea
\eeq
defined on the characteristic curves described by equations (\ref{charact1}b). If, in particular,
$\rho=0$, then the characteristic curves become straight lines and  equations (\ref{charact1}) are
integrated in the form:
\beq
\label{charact2}
\ba{l}
u=f(\eta_1,..,\eta_N), \\
x_k=\rho^{(k)}(u)t+\eta_k,~k=1,..,n,
\ea
\eeq
where
$\eta_k,~k=1,..,N$ are arbitrary constants, $f:\RR^n\to\RR$ is an
arbitrary scalar function that can be identified with the initial
condition: $f(x_1,..,x_n)=u(x_1,..,x_n,0)$ of the Cauchy problem in
$\RR^n$, and $u$ is constant along the characteristic straight
lines. From equations (\ref{charact2}) it also follows that the
general solution $u(x,t)$ of (\ref{equ-u}) for $\rho=0$ is defined
implicitly by the non-differential equation \beq \label{sol-u}
u=f\left(x_1-\rho^{(1)}(u)t,..,x_n-\rho^{(n)}(u)t\right). \eeq

Vector generalizations of equation (\ref{equ-u}) have been
considered in several papers \cite{ts,dn,ts2,F,K1,K2}.

The main result of our work is the construction, using elementary
spectral means, of the general solution of the following matrix
generalization of equation (\ref{equ-u}):
\begin{eqnarray}
\label{equ-w}
w_t+\sum\limits_{k=1}^nw_{x_k}\rho^{(k)}(w)=\rho(w)+
[w,T \tilde \rho(w)],
\end{eqnarray}
where $w$ is the unknown $N\times N$ matrix function of the $n+1$
independent variables $(x_1,..,x_n,t)\in\RR^{n+1}$, $T$ is any
constant diagonal matrix, $[\cdot ,\cdot ]$ is the usual commutator
between matrices, and $\rho^{(k)},~\rho,~\tilde
\rho:~\RR\to\RR,~k=1,..,n$ are $n+2$ arbitrary scalar functions
representable by positive power series, so that the quantities
$\rho^{(k)}(w),~k=1,..,n$, $\rho(w)$ and $\tilde\rho(w)$ are
well-defined functions of the matrix $w$.

In \S 2 we show that, as $w$ evolves according to (\ref{equ-w}), the
evolution of its spectrum decouples, and it is ruled by the scalar
analogue of the above equation. Therefore the eigenvalues and
suitably normalized eigenvectors of $w$ are the $N^2$ Riemann
invariants. In addition, the number of characteristic curves is only
$N$ (and not $N^2$, as it would be for a more generic vector system
of $N^2$ equations).

In \S 3 we construct, in the case $\rho=\tilde\rho=0$, the matrix
generalization of the non-differential equation (\ref{sol-u}), given
by a nonlinear system of $N^2$ non-differential equations for the
components of $w$.

In \S 4 we introduce a distinguished reduction of equation
(\ref{equ-w}) to a system of $N$ equations exhibiting, as Riemann
invariants, the eigenvalues of $w$.

We end this introduction with the obvious remark that solving
equation (\ref{equ-w}) is equivalent to solving the equation
transposed to (\ref{equ-w})
\beq
{\tilde
w}_t+\sum\limits_{k=1}^n\rho^{(k)}({\tilde w}){\tilde
w}_{x_k}=\rho(\tilde w)+[\tilde \rho(\tilde w)T,\tilde w],
\eeq
whose solution is just the transposed of $w$: $\tilde w=w^T$.%

\section{The spectral solution of (\ref{equ-w})}

The main characterization of the matrix equation (\ref{equ-w}) is
that, although the $N^2$ components $w_{ij},~i,j=1,..,N$ of $w$ are
coupled by the evolution, such a coupling disappears in spectral
space, and the eigenvalues of $w$ evolve according to the scalar
analogue (\ref{equ-u}) of (\ref{equ-w}).

\vskip 5pt \noindent {\bf Proposition 1} Let
$\{e^{(1)},..,e^{(N)}\}$ be the eigenvalues of matrix $w$, let
$\{\underline{v}^{(1)},..,\underline{v}^{(N)}\}$ be the
corresponding right eigenvectors, conveniently normalized to satisfy
the $N$ conditions ${v}^{(j)}_j=1,~j=1,..,N$, where ${v}^{(j)}_i$ is
the $i$-component of vector $\underline{v}^{(j)}$. Then, if $w$
evolves according to the matrix PDE (\ref{equ-w}), the dynamics in
spectral space is decoupled in the following way.

\vskip 5pt
\noindent
i) Each eigenvalue evolves separately according to the scalar analogue of equation (\ref{equ-w}):
\beq
\label{e-equ}
e^{(j)}_t+\sum\limits_{k=1}^ne^{(j)}_{x_k}\rho^{(k)}(e^{(j)})=\rho(e^{(j)}),~~j=1,..,N.
\eeq

\vskip 5pt \noindent ii) Known $e^{(j)}$ from equation
(\ref{e-equ}), the corresponding normalized eigenvector
$\underline{v}^{(j)}$ evolves according to the linear equation: \beq
\label{v-equ}
\left({v}^{(j)}_{i}\right)_t+\sum\limits_{k=1}^n\left({v}^{(j)}_{i}\right)_{x_k}\rho^{(k)}(e^{(j)})=
(T_j-T_i){v}^{(j)}_i\tilde\rho(e^{(j)}),~~i,j=1,..,N.
\eeq

\vskip 5pt \noindent Proof. The proof is by direct calculation. We
substitute the spectral decomposition of $w$: \beq
\label{reconstruction} w=v~E~v^{-1}, \eeq into (\ref{equ-w}), where
$E=\mbox{diag}(e^{(1)},..,e^{(N)})$ and $v$ is the matrix of the
normalized eigenvectors: $v_{ij}={v}^{(j)}_{i}$. After multiplying
this equation from the left and from the right respectively by
$v^{-1}$ and $v$, we obtain, after some manipulation: \beq
E_t+\sum\limits_{k=1}^nE_{x_k}\rho^{(k)}(E)-\rho(E)+
\left[v^{-1}\left(v_t+\sum\limits_{k=1}^nv_{x_k}\rho^{(k)}(E)
+Tv\tilde \rho(E)\right),E \right]=0. \eeq

Separating the diagonal and off-diagonal parts of this equation, we
obtain:
\beq \label{evol-Ev}
\ba{l}
E_t+\sum\limits_{k=1}^nE_{x_k}\rho^{(k)}(E)=\rho(E), \\
v_t+\sum\limits_{k=1}^nv_{x_k}\rho^{(k)}(E)+ Tv\tilde
\rho(E)=vh,~~h~\mbox{diagonal} \ea \eeq The diagonal matrix $h$ is
fixed by the normalization of matrix $v$. In our case, the diagonal
part of $v$ is the identity matrix; therefore, taking the diagonal
part of (\ref{evol-Ev}b), one infers that $h=T\tilde \rho(E)$.
Equations (\ref{evol-Ev}a) and (\ref{evol-Ev}b) with $h=T\tilde
\rho(E)$ are just equations (\ref{e-equ}) and (\ref{v-equ})
respectively. $\Box$

The PDEs (\ref{e-equ}) and (\ref{v-equ}) are converted into the
system of ODEs:
\beq \label{charact-e}
\ba{l}
\frac{de^{(j)}}{dt}=\rho(e^{(j)}),~~j=1,..,N \\
\frac{d{v}^{(j)}_i}{dt}=(T_j-T_i){v}^{(j)}_i\tilde\rho(e^{(j)}),~~i,j=1,..,N
\\ \frac{dx_k}{dt}=\rho^{(k)}(e^{(j)})~,k=1,..,n,
\ea
\eeq
defined on the characteristic curves described by equations %
(\ref{charact-e}c). On such characteristic curves, the eigenvalues
are obtained by the quadratures associated with (\ref{charact-e}a);
once the eigenvalues are constructed, the eigenvectors are obtained
solving the linear equations (\ref{charact-e}b).

If $\rho=\tilde\rho=0$, the characteristic curves become the
straight lines \beq \label{j-curve}
x_k=\rho^{(k)}(e^{(j)})t+\eta_k,~~k=1,..,n, \eeq where $\eta_k$ are
arbitrary integration constants, and each eigenvalue
$e^{(j)}(x_1,..,x_n,t)$ is defined by the implicit equation \beq
\label{e-sol} e^{(j)}=\epsilon^{(j)}\left(\eta_1,..,\eta_n\right)=
\epsilon^{(j)}\left(x_1-\rho^{(1)}(e^{(j)})t,..,x_n-\rho^{(n)}(e^{(j)})t\right),
\eeq where $\epsilon^{(j)}:\RR^n\to\RR$ is an arbitrary function.
For the normalized eigenvector $\underline{v}^{(j)}$, constant along
the characteristic curve (\ref{j-curve}), the  general solution is
explicit in terms of $e^{(j)}$:
\beq \label{v-sol}
\underline{v}^{(j)}(x_1,..,x_n,t)= \underline{\mbox{\bf
v}}^{(j)}\left(x_1-\rho^{(1)}(e^{(j)})t,..,x_n-\rho^{(n)}(e^{(j)})t\right),
\eeq
where ${\mbox{\bf v}}^{(j)}:\RR^n\to \RR^N$ is an arbitrary
vector function. Then the  general solution of the matrix system
(\ref{equ-w}) is achieved through the spectral formula (\ref
{reconstruction}). Such a solution depends, as it has to be, on the
$N^2$ arbitrary scalar functions $\epsilon^{(j)},~j=1,..,N$ and
${\mbox{\bf v}}^{(j)}_i,~i,j=1,..,N,~i\ne j$. We remark that, {\bf
although the evolution in spectral space is decoupled, the
components of $w$ evolve along {\bf all} the $N$ characteristics,
through  the algebraic coupling given by (\ref{reconstruction})}.

If, in particular, one is interested in solving the Cauchy problem
for the matrix system (\ref{equ-w}) on $\RR^n$,  for
$\rho=\tilde\rho=0$, one first remarks that, in this case,
$\epsilon^{(j)},~j=1,..,N$ and ${\mbox{\bf
v}}^{(j)}_i,~i,j=1,..,N,~i\ne j$ are identified with the initial
conditions for the eigenvalues and eigenvectors:
\beq \label{t=0}
\ba{l}
\epsilon^{(j)}(x_1,..,x_n)=e^{(j)}(x_1,..,x_n,0), \\
\underline{\mbox{\bf
v}}^{(j)}(x_1,..,x_n)=\underline{v}^{(j)}(x_1,..,x_n,0),
\ea
\eeq
Therefore: i) given the initial condition \beq \label{w0}
f(x_1,..,x_n)=w(x_1,..,x_n,0), \eeq one uniquely constructs the
initial conditions ${\epsilon}^{(j)}(x_1,..,x_n)$ and
$\underline{\mbox{\bf v}}^{(j)}(x_1,..,x_n)$ for the eigenvalues and
eigenvectors solving the associated eigenvalue problem:
\beq\label{fv=ve}
 f\underline{\mbox{\bf
v}}^{(j)}={\epsilon}^{(j)}\underline{\mbox{\bf v}}^{(j)},~~j=1,..,N
\eeq
with the above normalization for the eigenvectors. ii) Given
${\epsilon}^{(j)}$ and $\underline{\mbox{\bf v}}^{(j)}$, one obtains
the decoupled evolution of the spectrum through the equations
(\ref{e-sol}) and (\ref{v-sol}). iii) The reconstruction of
$w(x_1,..,x_n,t)$ is finally achieved through the formula
(\ref{reconstruction}).

The elementary spectral solution of the nonlinear PDE (\ref{equ-w}) presented in this section is,
to the best of our knowledge, new.

The nonlinear  PDEs (\ref{equ-w}) have been recently identified by
the authors because they arise as the simplest examples of a class of nonlinear systems of PDEs in arbitrary
dimensions generated by a novel dressing procedure based on a homogeneous integral equation
with nontrivial kernel \cite{ZS}.

\section{The analogue of the implicit equation (\ref{sol-u})}

Starting with the spectral solution of (\ref{equ-w}) obtained in the
previous section, it is possible to construct, if
$\rho=\tilde\rho=0$, the matrix analogue of the implicit
non-differential equation (\ref{sol-u}).

\vskip 5pt \noindent {\bf Proposition 2} Let
$f_{ij}:\RR^n\to\RR,~i,j=1,..,N$ be $N^2$ arbitrary scalar functions
admitting formal positive power expansions, so that
$f_{ij}(M_1,..,M_n)$ are well-defined $N\times N$ matrices, where
$M_1,..,M_n$ are arbitrary matrices $N\times N$. Denote by
$(f_{ij}(M_1,..,M_n))_{kl}$ the $(kl)$-component of the matrix
$f_{ij}(M_1,..,M_n)$. Then the general solution $w$ of equation
(\ref{equ-w}) is characterized implicitly by the following system of
$N^2$ non-differential equation: \beq \label{sol3}
w_{ij}=\sum\limits_{k=1}^N\left(f_{ik}\left(x_1I-\rho^{(1)}(w)t,..,x_nI-
\rho^{(n)}(w)t\right)\right)_{kj},~i,j=1,..,N, \eeq where $w_{ij}$
is the $(ij)$-component of matrix $w$ and $I$ is the $N\times N$
identity matrix.

\vskip 5pt
\noindent
Proof. Using equations (\ref{reconstruction}), (\ref{e-equ}) and (\ref{v-equ}), %
we have that:
\beq\label{w}
\ba{l}
w_{ij}=\sum_{k=1}^N v_{ik} e^{(k)}
(v^{-1})_{kj}=
 \sum_{l=1}^N {\mbox{\bf{v}}}_{ik}(x_1-\rho^{(1)}(e^{(k)}) t,\dots,
 x_n-\rho^{(n)}(e^{(k)}) t) \times \\
 \epsilon^{(k)}(x_1-\rho^{(1)}(e^{(k)}) t ,\dots,
 x_n-\rho^{(n)}(e^{(k)}) t)(\mbox{\bf{v}}^{-1})_{kj}
\ea
\eeq
Let $f$ be the matrix having $\epsilon^{(l)},l=1,\dots,N$ %
as eigenvalues and ${\mbox{\bf{v}}}$ as matrix of eigenvectors,
i.e.: $\displaystyle f_{ij}=\sum_{l=1}^N {\mbox{\bf v}}_{il}
\epsilon ^{(l)} {\mbox{\bf v}}^{-1}_{lj}$, then equation (\ref{w})
becomes:
\beq\label{wbis}
w_{ij}=
\sum_{k,l=1}^N{\mbox{\bf{v}}}_{lk}(x_1-\rho^{(1)}(e^{(k)}) t ,\dots,
 x_n-\rho^{(n)}(e^{(k)}) t)f_{il}(x_1-\rho^{(1)}(e^{(k)}) t ,\dots,
 x_n-\rho^{(n)}(e^{(k)}) t)(\mbox{\bf{v}}^{-1})_{kj}.
\eeq
Replacing now the definition of function of a matrix: %
\beq
\ba{l}
f_{il}(x_1-\rho^{(1)}(e^{(k)}) t
,\dots,x_n-\rho^{(n)}(e^{(k)})t)\delta_{ks}=
\\
\sum\limits_{m,r=1}^N({\mbox{\bf{v}}}^{-1})_{km}\left(f_{il}(x_1I-\rho^{(1)}(w)
t ,\dots,x_nI-\rho^{(n)}(w)t)\right)_{mr}{\mbox{\bf{v}}}_{rs},
\ea
\eeq
into equation (\ref{wbis}), we obtain the result.$\Box$%

We end this section remarking that, as a byproduct of Propositions 1
and 2, the complicated system of $N^2$ non-differential equations
(\ref{sol3}) defining the solution $w$ is reduced to the solution of
the single non-differential equation (\ref{e-sol}) for the
eigenvalues of $w$.

\section{A distinguished reduction}

In this section we briefly describe a reduction of the matrix equation (\ref{equ-w})
exhibiting Riemann invariants coinciding with the eigenvalues $e^{(j)},~j=1,..,N$.

Consider the subspace of $N\times N$ matrices spanned by the basis
$\{\omega_0,..,\omega_{N-1}\}$ given by:
\beq \label{basis}
\omega_0=I,~\left(\omega_1\right)_{ij}=\delta_{i+1,j},~..~,~\left(\omega_k\right)_{ij}=\delta_{i+k,j},~..~,
~\left(\omega_{N-1}\right)_{ij}=\delta_{i+N-1,j},~~~\mbox{mod} N.
\eeq
This subspace is left invariant under matrix multiplication,
since: %
\beq
\omega_j\omega_k=\omega_k\omega_j=\omega_{j+k},~~~ \mbox{mod}
N,
\eeq
therefore it defines a reduction of (\ref{equ-w}) from the %
$N^2$ components of $w$ to the $N$ scalar coefficients
$\nu_k,~k=1,..,N$ of the expansion: \beq \label{reduction}
w=\sum\limits_{k=1}^{N}\nu_k\omega_{k-1}. \eeq

We remark that the mapping between the $N$ dependent variables
$\nu_k,~k=1,..,N$ of the reduced system  and the eigenvalues
$e^{(j)},~j=1,..,N$ allows one to {\bf decouple completely} the
above dynamics; therefore the eigenvalues $e^{(j)},~j=1,..,N$ are
the Riemann invariants for this reduced class of equations.

In the remaining part of this section we show some examples of
reduced systems.

A system of $2$ interacting fields in $2+1$ dimensions is obtained,
for instance, choosing
$N=2,~n=2,~\rho^{(1)}(x)=x,~\rho^{(2)}(x)=ax^2,~\rho=\tilde\rho=0$:
\beq\label{2+1} \ba{l}
{\nu_1}_t+\nu_1{\nu_1}_{x_1}+\nu_2{\nu_2}_{x_1}+a(\nu^2_1+\nu^2_2){\nu_1}_{x_2}+2a\nu_1\nu_2{\nu_2}_{x_2}=0, \\
{\nu_2}_t+\nu_2{\nu_1}_{x_1}+\nu_1{\nu_2}_{x_1}+2a\nu_1\nu_2{\nu_1}_{x_2}+a(\nu^2_1+\nu^2_2){\nu_2}_{x_2}=0.
\ea \eeq
If $a=0$, equation (\ref{2+1}) reduces to a particular case of the %
well-known gas dynamics equations \cite{Whitham}: \beq \label{gas1}
\ba{l}
{\nu_1}_t+\nu_1{\nu_1}_{x_1}+\nu_2{\nu_2}_{x_1}=0, \\
{\nu_2}_t+\nu_2{\nu_1}_{x_1}+\nu_1{\nu_2}_{x_1}=0.
\ea
\eeq
The eigenvalues of $w$: %
\beq
 e^{(1)}=\nu_1+\nu_2,~~~~~e^{(2)}=\nu_1-\nu_2,
\eeq
evolve decoupled according to equation (\ref{e-equ}) and are the Riemann invariants %
of the above two systems (\ref{2+1}) and (\ref{gas1}).

The simplest example of a system of $3$ interacting fields arises
choosing $N=3,~n=1,~\rho^{(1)}(x)=x,~\rho=\tilde\rho=0$:
\beq
\label{gas3}
\ba{l}
{\nu_1}_t+\nu_1{\nu_1}_{x_1}+\nu_3{\nu_2}_{x_1}+\nu_2{\nu_3}_{x_1}=0, \\
{\nu_2}_t+\nu_2{\nu_1}_{x_1}+\nu_1{\nu_2}_{x_1}+\nu_3{\nu_3}_{x_1}=0, \\
{\nu_3}_t+\nu_3{\nu_1}_{x_1}+\nu_2{\nu_2}_{x_1}+\nu_1{\nu_3}_{x_1}=0.
\ea
\eeq
The three eigenvalues %
\beq
\ba{l}
 e^{(1)}=\nu_1+\nu_2+\nu_3,~~~~~e^{(2)}=\frac{1}{2}\left[2\nu_1-\nu_2-\nu_3+i\sqrt{3}|\nu_2-\nu_3|\right], \\
 e^{(3)}=\frac{1}{2}\left[2\nu_1-\nu_2-\nu_3-i\sqrt{3}|\nu_2-\nu_3|\right]
\ea
\eeq
evolve decoupled according to equation (\ref{e-equ}). But only $e^{(1)}$ is real, therefore %
the system is not hyperbolic and the eigenvalues $e^{(2)}$ and
$e^{(3)}$ can be called Riemann invariants only in an extended
sense.

The systematic study of this reduced class and of other reductions
of equation (\ref{equ-w}), in the search for integrable and
applicative equations, is postponed to a subsequent paper.

\vskip 10pt \noindent {\bf Acknowledgments}. PMS was supported by
the INFN grant 2006 and by the Centro Internacional de Ciencias
(CIC) of Cuernavaca (Mexico), where this manuscript was partially
written. AIZ was supported by the INTAS Young Scientists Fellowship
Nr. 04-83-2983, by the RFBR grants 04-01-00508, 06-01-90840,
06-01-92053 and by
the grant Ns 7550.2006.2. PMS acknowledges useful discussions with
R. Beals and F. Calogero. AIZ acknowledges useful discussions with
A. Maltsev, M. Pavlov and E. Ferapontov.


\end{document}